\documentclass[final,5p,times,twocolumn,nonatbib]{elsarticle}

\usepackage[T1]{fontenc}
\usepackage[utf8]{inputenc}
\usepackage{amsmath}
\usepackage{booktabs}
\usepackage{graphicx}
\usepackage{siunitx}
\usepackage{tabularx}
\usepackage[hidelinks]{hyperref}
\usepackage[
  backend=biber,
  style=numeric-comp,
  sorting=none,
  maxbibnames=99,
  giveninits=true
]{biblatex}

\graphicspath{{figures/}}
\DeclareSIUnit{\angstromunit}{\text{\AA}}
\DeclareSIUnit{\rydbergunit}{Ry}
\DeclareSIUnit{\hartreeunit}{Ha}
\DeclareSIUnit{\bohrunit}{bohr}
\DeclareSIUnit{\elementarychargeunit}{\ensuremath{e}}
\begin{document}

\begin{frontmatter}

\title{C 1s core-level fingerprints of reconstructed titanium vacancies in titanium carbide}

\author[muni]{Jakub Koch}
\author[muni]{Pavel Ondra{\v{c}}ka\corref{corresponding}}
\ead{ondracka@mail.muni.cz}
\cortext[corresponding]{Corresponding author.}

\address[muni]{Department of Plasma Physics and Technology, Faculty of Science, Masaryk University, Kotl{\'a}{\v{r}}sk{\'a} 2, CZ-61137 Brno, Czech Republic}

\begin{abstract}
Recent first-principles searches predict that a titanium vacancy in rocksalt TiC is not an empty octahedral site but can reconstruct through the formation of carbon Frenkel pairs and short C--C bonds. We determine the C 1s signatures of these local structures and assess whether they remain specific in the presence of disordered carbon. Site-resolved binding energies were calculated with a self-consistent core-hole method for an unreconstructed vacancy and four representative C--C-bonded reconstructions. The unreconstructed vacancy shifts its nearest-neighbour C 1s level by only \SI{-0.36}{\electronvolt}. In contrast, the largest positive shift in each reconstructed model ranges from \SI{+2.34}{\electronvolt} to \SI{+2.80}{\electronvolt}. The large shifts occur at carbon atoms participating in the reconstructed C--C network and correlate with markedly less negative Mulliken charges, providing candidate local fingerprints of vacancy reconstruction. However, an exploratory TiC/amorphous-C model shows that disordered carbon environments can extend into the same high-binding-energy range. Overall, the calculated shifts provide reference energies for experimental searches for reconstructed Ti vacancies in TiC.
\end{abstract}


\end{frontmatter}

\section{Introduction}

Titanium carbide is a refractory, electrically conducting ceramic used in wear-resistant coatings and cutting tools \cite{Rizzo2020}. Its combination of hardness and conductivity has also motivated TiC-containing electrical-contact materials, including coatings of nanocrystalline TiC in an amorphous-carbon matrix (nc-TiC/a-C) and Ag--TiC switching contacts \cite{Lewin2006,Liu2026Contacts}. The mechanical and electronic properties of TiC are sensitive to composition and point defects \cite{Guemmaz2000}. Its rocksalt (B1) lattice accommodates substantial non-stoichiometry, predominantly through carbon vacancies, which can be abundant over its homogeneity range \cite{Tsetseris2008,GreczynskiCarbides2018}. Although titanium vacancies are expected to be much less common under many conditions, they are central to proposed mechanisms of metal self-diffusion and therefore relevant to high-temperature evolution \cite{Tsetseris2008,Tang2020}.

The conventional model treats a titanium vacancy as an empty metal site surrounded by a slightly relaxed carbon octahedron. Candidate models of Ti self-diffusion include Ti vacancies associated with several C vacancies and competing off-lattice configurations \cite{Tang2020,Salehin2021}. Smirnova, Nourazar, and Korzhavyi identified another defect family in which the carbon neighbours of a single Ti vacancy rearrange internally into lower-energy C--C-bonded configurations \cite{Smirnova2024}. Their systematic search found several such minima. The planar-dicarbon configuration, containing a carbon dimer connected to four surrounding carbon atoms, is \SI{3.5}{\electronvolt} below the unreconstructed vacancy and was identified as the ground-state geometry in TiC. Other metastable structures contain different C--C networks and leave different amounts of open volume for metal motion.

The relevance of this reconstruction extends beyond the isolated vacancy. For equiatomic TiC, ZrC, and HfC, calculations including the reconstructed metal vacancy favour dissociated Schottky pairs and reverse the nearest-neighbour metal-vacancy/C-vacancy interaction from attraction to repulsion \cite{Nourazar2026}. Nourazar and Korzhavyi attribute the strong repulsion to the loss of one of the five C--C bonds stabilizing the planar-dicarbon structure. Jiang et al. independently recovered the same Ti-vacancy geometry, describing it as a dicarbon antisite bound to the two C vacancies left by the displaced carbon atoms \cite{Jiang2025}. They distinguish this reconstructed vacancy from an isolated dicarbon antisite without accompanying C vacancies, and their thermodynamic model predicts composition- and temperature-dependent populations of these and other defects. Their cluster calculations also find favourable binding for a Ti vacancy associated with four to six C vacancies, with the reconstructed Ti vacancy as the reference \cite{Jiang2025}. Both internal reconstruction and C-vacancy association therefore remain relevant to the defect populations.

These developments reinforce the need for an experimental observable that responds to the internal vacancy geometry. To our knowledge, however, no experiment has yet assigned one of the reconstructed Ti-vacancy configurations directly.
Core-level photoelectron spectroscopy is a natural candidate because the binding energy (BE) of a core electron depends on its local chemical and electrostatic environment \cite{Bagus1999}. First-principles core-level calculations have helped distinguish implanted nitrogen configurations in TiO$_2$ \cite{Panepinto2020} and relate vacancy-induced Mo 3d and S 2p shifts to sulfur-vacancy formation in MoS$_2$ \cite{Ozaki2023}. In Ti-based ceramics, the same strategy has identified Ti-vacancy and N-interstitial fingerprints and, with statistical modelling and independent compositional constraints, enabled Ti-vacancy quantification in titanium oxynitride and Ti-deficient TiN \cite{Ondracka2022,Ondracka2025}.

Likewise, calculations can reject seemingly plausible spectral assignments. A common example is the attribution of high-BE O 1s components in metal oxides to oxygen vacancies, and the use of their fitted areas for vacancy quantification \cite{Wang2024,Easton2025}. In typical metal oxides, O atoms bond to cations rather than other O atoms, so vacancy-induced shifts of the neighbouring cation core levels offer an alternative probe if sufficiently large. First-principles studies of ZnO found no distinct component from the remaining O atoms near bulk vacancies and instead identified interstitial or surface species as plausible origins of the high-BE intensity \cite{Frankcombe2023,Li2023}. Collectively, these examples define the conditions that a useful point-defect fingerprint must satisfy: a resolvable shift, enough contributing atoms, a sufficiently large defect population, and limited overlap with unrelated chemical environments.

Spectral overlap is a particular concern in the C 1s region of TiC. Nanocomposite TiC coatings can contain nanocrystalline TiC embedded in an amorphous-carbon tissue phase \cite{Lewin2006,Soucek2017}, while air-exposed surfaces introduce additional non-carbide carbon contributions and BE-referencing complications \cite{GreczynskiCarbides2018,GreczynskiCarbon2018}. Although adventitious carbon remains a pervasive complication in XPS, low-energy Ar gas-cluster ion-beam sputtering can preferentially remove organic overlayers from inorganic surfaces with less damage than conventional monoatomic Ar$^+$ sputtering, provided that the irradiation conditions are validated for the substrate \cite{Miisho2017,Shard2024}. Carbonaceous environments span a wide range of local bonding motifs and C 1s BEs \cite{Aarva2019}, and metal chemistry can alter the structure of the amorphous-carbon matrix \cite{Endrino2024}. The same C--C bonds that make a reconstructed vacancy spectroscopically distinctive in an ideal crystal may therefore place its signal in a congested part of an experimental spectrum.

Here we calculate site-resolved C 1s BEs for an unreconstructed Ti vacancy and four representative C--C-bonded reconstructions based on the motifs identified by Smirnova et al. \cite{Smirnova2024}. We ask three deliberately separate questions: (i) which local sites exhibit a sizeable BE shift, (ii) whether the BE shifts distinguish reconstruction from an unreconstructed vacancy or distinguish the individual reconstructions from one another, and (iii) whether the predicted high-BE sites remain specific in the presence of amorphous carbon.

\section{Computational methodology}

We used the same fully relaxed atomic structures reported in the configurational search of Smirnova et al. \cite{Smirnova2024}. Each model is a $3\times3\times3$ repetition of the eight-site conventional B1 cell with a lattice parameter of \SI{4.33}{\angstromunit}, containing 107 Ti atoms, 108 C atoms, and one vacant Ti site. We examined the unreconstructed Ti vacancy, denoted 0, together with reconstructed minima 1A, 2G, 2I, and 3C.

\begin{figure*}[!t]
\centering
\includegraphics[width=0.98\textwidth]{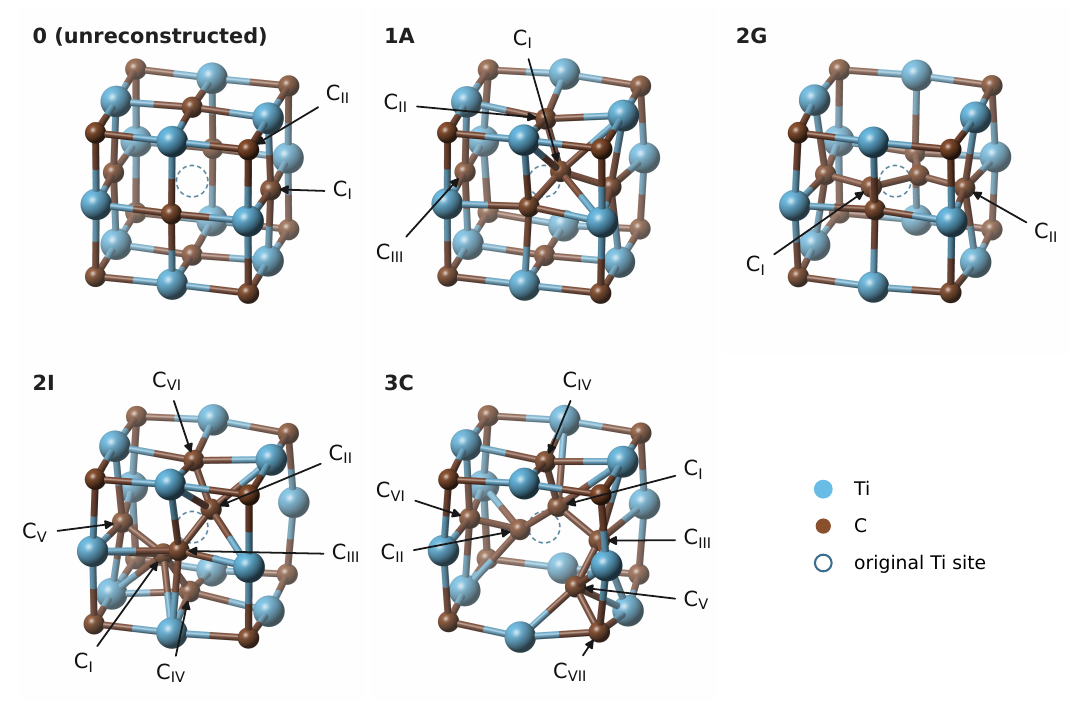}
\caption{Vacancy-centred views of the unreconstructed vacancy (0) and the reconstructed Ti-vacancy models. Ti and C atoms are light blue and brown, respectively, and the dashed circle marks the original Ti site. Labels identify the marked carbon sites listed in Table~\ref{tab:sites} and are ordered by distance from the original Ti site. The displayed bonds are included only as visual guides.}
\label{fig:structures}
\end{figure*}

The four reconstructions were selected to span distinct low-energy motifs: the lowest-energy one-Frenkel-pair minimum (1A), the overall ground state (2G), the lowest of the asymmetric two-pair structures that retain an open vacancy centre (2I), and an energetically favourable three-pair topology (3C). For 3C, we follow the branched seven-carbon geometry illustrated in Fig.~2(e) of Ref.~\cite{Smirnova2024}. Table~\ref{tab:structures} summarizes the published reconstruction energies and local motifs, and Fig.~\ref{fig:structures} shows the corresponding geometries and carbon environments analysed below.

\begin{table}[htbp]
\centering
\small
\caption{Vacancy configurations considered in this work, following the nomenclature of Smirnova et al. \cite{Smirnova2024}. Reconstruction energies are taken from that work and are relative to the unreconstructed vacancy (0).}
\label{tab:structures}
\begin{tabularx}{\columnwidth}{@{}lc>{\raggedright\arraybackslash}X@{}}
\toprule
Structure & $\Delta E_\mathrm{rec}$ (\si{\electronvolt}) & Local carbon motif \\
\midrule
0 (unreconstructed) & 0.0 & relaxed C octahedron with no new C--C bond \\
1A & $-1.9$ & approximately planar three-bond reconstruction around one displaced C \\
2G & $-3.5$ & planar-dicarbon reconstruction connected to four lattice C atoms \\
2I & $-2.0$ & asymmetric five-bond network formed by two displaced C atoms \\
3C & $\approx-0.5$ & branched seven-carbon network formed by three displaced C atoms \\
\bottomrule
\end{tabularx}
\end{table}

Single-point ground-state and core-hole calculations were performed with OpenMX 3.9.9 \cite{Ozaki2003} using the PBE generalized-gradient approximation \cite{Perdew1996} and the DFT\_DATA19 pseudopotential release. The localized basis sets were \texttt{C6.0\_1s-s3p2d1} and \texttt{Ti7.0-s3p2d1}. Empty basis centres were retained at the vacant Ti and C lattice sites.

The crystalline calculations were spin-polarized and used a \SI{400}{\rydbergunit} real-space integration cutoff, a $3\times3\times3$ Monkhorst--Pack mesh, an electronic temperature of \SI{300}{\kelvin}, and a self-consistency criterion of \SI{1e-5}{\hartreeunit}. A full C 1s core hole was imposed on one selected carbon at a time. The electron removed from the core level was placed at the Fermi level, so the initial and final states are neutral and contain $N$ electrons. The exact Coulomb-cutoff treatment was enabled to suppress interactions between periodically repeated core holes \cite{OzakiLee2017}. The final state was calculated self-consistently, including electronic relaxation and screening of the core hole.

For site $i$, $E_\mathrm{initial}(N)$ denotes the normal DFT ground-state energy and $E_{\mathrm{final},i}(N)$ the energy of the self-consistent final state with a C 1s core hole on that site. The C 1s BE was evaluated as:
\begin{equation}
E_\mathrm{B}(i)=E_{\mathrm{final},i}(N)-E_\mathrm{initial}(N).
\label{eq:binding}
\end{equation}

The crystalline chemical shift was referenced within each structure to a bulk-like carbon site at a minimum-image distance of approximately \SI{11.2}{\angstromunit} from the vacancy,
\begin{equation}
\Delta E_\mathrm{B}(i)=E_\mathrm{B}(i)-E_\mathrm{B}(\mathrm{ref}).
\label{eq:shift}
\end{equation}
Relative BE shifts are emphasized because systematic and numerical errors tend to cancel when taking differences between sites \cite{OzakiLee2017,Ondracka2022}. The same approach nevertheless gives useful absolute values. Across the eight carbon-containing molecules and solids in the OpenMX PBE19\_1s benchmark, the mean absolute error is \SI{0.34}{\electronvolt}, including a calculated TiC C 1s BE of \SI{281.43}{\electronvolt} compared with the listed experimental value of \SI{281.5}{\electronvolt} \cite{OpenMXCarbonBenchmark}. Convergence tests on the 215-atom 2G TiC model showed that the relative C 1s BE shifts of two representative carbon environments are converged within approximately \SI{10}{\percent} with respect to the real-space grid cutoff, $k$-point density, and basis-set size. Mulliken charges were read from the ground-state calculation and serve as basis-dependent descriptors of charge trends \cite{Bagus1999}.
For the multiplicity analysis, Ti--C and short C--C neighbours were identified within \SI{2.60}{\angstromunit} and \SI{1.75}{\angstromunit}, respectively.

Because nanocrystalline TiC coatings commonly contain an amorphous-carbon tissue phase \cite{Lewin2006,Soucek2017}, we constructed a periodic a-C/TiC(001) slab model to test spectral overlap with disordered carbon. An amorphous-carbon precursor at an initial density of approximately \SI{2.0}{\gram\per\centi\metre\cubed} was generated by spin-unpolarized Born--Oppenheimer molecular dynamics. The preparation stage used a \SI{250}{\rydbergunit} cutoff, a $2\times2\times4$ mesh, an electronic temperature of \SI{1000}{\kelvin}, a \SI{2}{\femto\second} time step, and an NVT Nos{\'e}--Hoover thermostat. The system was held at \SI{2500}{\kelvin} for 500 steps and cooled linearly to \SI{300}{\kelvin} by step 1500. An amorphous-C region from the quenched precursor was then joined to the 215-atom 2G TiC(001) slab. The resulting 332-atom periodic cell (107 Ti and 225 C) contained two TiC/amorphous-C interfaces and had dimensions of \numproduct{12.99x12.99x19.98}\,\si{\angstromunit}. This combined model was relaxed with a maximum force criterion of \SI{3e-4}{\hartreeunit\per\bohrunit}.

The parent and core-hole calculations for the final interface model followed the crystalline protocol, using a \SI{300}{\rydbergunit} real-space cutoff and a consistent $k$-point density with a $3\times3\times2$ mesh. In total, 143 C sites were sampled: 48 in the crystalline part (including the 2G reconstruction and interface-affected sites) and 95 in the amorphous part. Most sites within each region were chosen approximately at random, with deliberate inclusion of the 2G reconstruction and selected interface environments. The resulting site-resolved distribution was used to compare the energy ranges represented by the sampled crystalline and amorphous environments.

The data-extraction and visualization scripts were developed with assistance from OpenAI Codex and are available as a reproducible analysis workflow \cite{OndrackaWorkflow2026}.

\section{Results and discussion}

Table~\ref{tab:sites} summarizes the calculated C 1s BEs and local descriptors for the crystalline carbon sites marked in Fig.~\ref{fig:structures}. We report only sites with $\lvert\Delta E_{\mathrm{B}}\rvert\geq\SI{0.10}{\electronvolt}$, focusing on the most pronounced shifts. The labelled carbon atoms are ordered by distance from the vacant Ti site, and $m$ is the number of atoms represented by each entry per Ti vacancy. Integer values of $m$ denote symmetry-equivalent atoms, whereas $\approx m$ groups atoms with the same first-neighbour coordination but small differences caused by more distant relaxations. $E_{\mathrm{B}}$ is the calculated $\Delta$SCF BE, and $\Delta E_{\mathrm{B}}$ is referenced to C$_{\mathrm{ref}}$ within the same structure.

The unreconstructed vacancy produces only negative shifts among the retained sites, with the largest magnitude, \SI{-0.36}{\electronvolt}, at its six nearest-neighbour C atoms.
Reconstruction changes the response qualitatively. Each reconstructed structure has at least one carbon with a positive BE shift above \SI{2.3}{\electronvolt}. The largest values are \SI{+2.45}{\electronvolt} for 1A, \SI{+2.75}{\electronvolt} for 2G, \SI{+2.80}{\electronvolt} for 2I, and \SI{+2.34}{\electronvolt} for 3C. These maxima occur on displaced carbon atoms inside or near the Ti-vacancy volume.

\begin{table}[htbp]
\centering
\small
\caption{Calculated C 1s BEs and BE shifts for the affected carbon sites in the five crystalline vacancy models. $q$ is the ground-state Mulliken charge, and $r_{\mathrm{vac}}$ is the minimum-image distance from the vacant Ti site.}
\label{tab:sites}
\begingroup
\setlength{\tabcolsep}{4pt}
\renewcommand{\arraystretch}{0.98}
\newcommand{\CrystallineSiteRows}{%
\multicolumn{6}{@{}l}{\textbf{Unreconstructed vacancy (0)}} \\
C$_{\mathrm{I}}$ & 281.28 & $-0.36$ & $-0.71$ & 2.22 & 6 \\
C$_{\mathrm{II}}$ & 281.53 & $-0.11$ & $-0.77$ & 3.73 & 8 \\
\addlinespace[0.35em]
\multicolumn{6}{@{}l}{\textbf{1A}} \\
C$_{\mathrm{I}}$ & 284.03 & $+2.45$ & $-0.20$ & 1.04 & 1 \\
C$_{\mathrm{II}}$ & 282.03 & $+0.44$ & $-0.62$ & 1.89 & $\approx3$ \\
C$_{\mathrm{III}}$ & 281.23 & $-0.36$ & $-0.68$ & 2.24 & 2 \\
\addlinespace[0.35em]
\multicolumn{6}{@{}l}{\textbf{2G}} \\
C$_{\mathrm{I}}$ & 284.33 & $+2.75$ & $-0.20$ & 0.75 & 2 \\
C$_{\mathrm{II}}$ & 282.04 & $+0.45$ & $-0.66$ & 1.95 & 4 \\
\addlinespace[0.35em]
\multicolumn{6}{@{}l}{\textbf{2I}} \\
C$_{\mathrm{I}}$ & 284.39 & $+2.80$ & $-0.24$ & 1.08 & 1 \\
C$_{\mathrm{II}}$ & 283.47 & $+1.88$ & $-0.40$ & 1.32 & 1 \\
C$_{\mathrm{III}}$ & 283.19 & $+1.60$ & $-0.54$ & 1.58 & 1 \\
C$_{\mathrm{IV}}$ & 282.22 & $+0.62$ & $-0.63$ & 1.95 & 1 \\
C$_{\mathrm{V}}$ & 282.17 & $+0.58$ & $-0.63$ & 1.97 & 1 \\
C$_{\mathrm{VI}}$ & 282.30 & $+0.70$ & $-0.61$ & 2.02 & 1 \\
\addlinespace[0.35em]
\multicolumn{6}{@{}l}{\textbf{3C}} \\
C$_{\mathrm{I}}$ & 283.97 & $+2.34$ & $-0.21$ & 0.69 & 1 \\
C$_{\mathrm{II}}$ & 283.03 & $+1.40$ & $-0.32$ & 1.01 & 1 \\
C$_{\mathrm{III}}$ & 283.25 & $+1.62$ & $-0.51$ & 1.62 & 1 \\
C$_{\mathrm{IV}}$ & 282.13 & $+0.50$ & $-0.66$ & 1.97 & 1 \\
C$_{\mathrm{V}}$ & 282.98 & $+1.35$ & $-0.57$ & 2.03 & 1 \\
C$_{\mathrm{VI}}$ & 282.18 & $+0.55$ & $-0.66$ & 2.11 & 1 \\
C$_{\mathrm{VII}}$ & 282.69 & $+1.06$ & $-0.67$ & 3.53 & 1 \\
}

\begin{tabular}{@{}lrrrrr@{}}
\toprule
Site & $E_{\mathrm{B}}$ & $\Delta E_{\mathrm{B}}$ & $q$ & $r_{\mathrm{vac}}$ & $m$ \\
 & [\si{\electronvolt}] & [\si{\electronvolt}] & [\si{\elementarychargeunit}] & [\si{\angstromunit}] & \\
\midrule
\CrystallineSiteRows
\bottomrule
\end{tabular}
\endgroup
\end{table}

\begin{figure}[htbp]
\centering
\includegraphics[width=\linewidth]{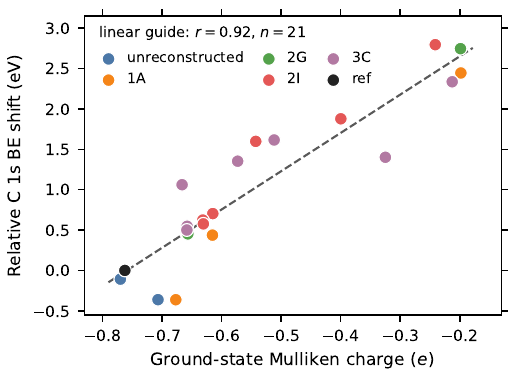}
\caption{Ground-state Mulliken charge versus C 1s BE shift for the carbon sites in Table~\ref{tab:sites} and the bulk-like, zero-shift TiC reference (``ref''). The dashed line is a least-squares fit.}
\label{fig:charge}
\end{figure}

The most strongly shifted sites are also markedly less negative than the bulk-like reference, whose Mulliken charge is approximately \SI{-0.76}{\elementarychargeunit} in every structure. For the sites shown in Fig.~\ref{fig:charge}, the Pearson correlation coefficient between the ground-state Mulliken charge and C 1s BE shift is 0.92. The correlation indicates that the largest positive shifts accompany charge redistribution when carbon changes from a predominantly Ti-coordinated lattice environment to participation in the reconstructed C--C network.

The negative BE shifts of undercoordinated but otherwise lattice-like C sites reveal a complementary electrostatic trend. In the unreconstructed vacancy and 1A, unbonded C atoms that remain close to their lattice positions but have five rather than six Ti neighbours shift by \SI{-0.36}{\electronvolt}. These atoms are \SIrange{0.06}{0.09}{\elementarychargeunit} less negative than the bulk-like reference, opposite to the overall charge--shift trend. Removal of a neighbouring, positively charged Ti atom reduces the stabilizing local electrostatic potential at C and can therefore contribute a negative BE shift. This response is similar in sign but smaller in magnitude than the \SI{-0.54}{\electronvolt} N 1s BE shift calculated for a Ti vacancy in TiN and the \SIrange{0.5}{0.6}{\electronvolt} decrease per neighbouring Ti vacancy found in titanium oxynitride \cite{Ondracka2022,Ondracka2025}. The smaller TiC response can be explained by the greater covalent mixing and lower ionic character of Ti--C relative to Ti--N bonding \cite{Didziulis1994}, resulting in a weaker local electrostatic (Madelung-like) response to the loss of a Ti neighbour.

The energetic separation defines what an ideal crystalline spectrum could reveal. The \SI{-0.36}{\electronvolt} nearest-neighbour BE shift of the unreconstructed vacancy is small relative to the spectral complexity normally encountered in TiC and would likely appear as unresolved asymmetry or broadening unless the vacancy fraction is substantial. By contrast, the high-BE sites of reconstructed vacancies could generate a shoulder or component in a sufficiently crystalline sample with an enhanced reconstructed-vacancy population, thereby distinguishing a C--C-bonded reconstruction from the conventional empty-site model.

Peak position alone cannot identify the individual reconstruction. The largest BE shifts of all four reconstructed models lie within \SI{0.46}{\electronvolt} of one another, and each structure also contributes sites at intermediate or near-bulk energies. Experimental broadening would merge several of these contributions. The multiplicities in Table~\ref{tab:sites} show that the most strongly shifted site occurs once per vacancy in 1A, 2I, and 3C and twice in 2G.

\begin{figure}[htbp]
\centering
\includegraphics[width=\linewidth]{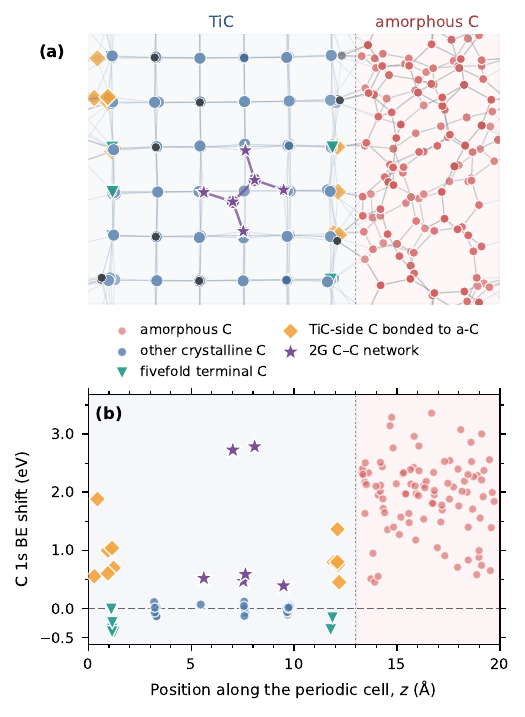}
\caption{Structure and C 1s BE shifts in the periodic 2G TiC/amorphous-C model. (a) Atomic structure with symbols marking the sampled C-site categories. (b) BE shifts relative to the \SI{281.58}{\electronvolt} median of 22 sampled bulk-like TiC-interior sites.}
\label{fig:interface}
\end{figure}

The TiC/amorphous-C model is shown in Fig.~\ref{fig:interface}(a), with the corresponding site-resolved C~1s BE shifts in Fig.~\ref{fig:interface}(b). The two central C atoms of the 2G vacancy reconstruction have BE shifts of \SI{+2.72}{\electronvolt} and \SI{+2.78}{\electronvolt} relative to the TiC interior. These values differ from the \SI{+2.75}{\electronvolt} shift in crystalline 2G by less than \SI{0.04}{\electronvolt}, consistent with the atoms' inequivalent positions relative to the interface. The four neighbouring C$_{\mathrm{II}}$ sites have shifts of \SIrange{+0.39}{+0.59}{\electronvolt}, with a mean of \SI{+0.49}{\electronvolt}, compared with \SI{+0.45}{\electronvolt} in crystalline 2G.

The sampled sites in the amorphous region have a mean BE shift of \SI{+1.94}{\electronvolt} and a median of \SI{+2.02}{\electronvolt}, spanning \SIrange{+0.45}{+3.36}{\electronvolt}.
C atoms in the terminating TiC(001) planes form two distinct groups. The 11 sampled atoms bonded directly to the amorphous network have a mean BE shift of \SI{+0.90}{\electronvolt}, intermediate between bulk-like TiC and amorphous carbon. In contrast, the nine fivefold-coordinated atoms without short C--C contacts have a mean shift of \SI{-0.30}{\electronvolt} (median \SI{-0.36}{\electronvolt}), reproducing the undercoordination response in the crystalline vacancy models. Their mean Mulliken charge is \SI{-0.68}{\elementarychargeunit}, compared with \SI{-0.77}{\elementarychargeunit} for the interior reference group, again giving the opposite sign to a charge-only expectation. The opposite signs of the two groups are consistent with direct C--C bonding and the associated electronic redistribution overriding the negative electrostatic contribution from reduced Ti coordination.

The two central 2G sites lie within the amorphous-carbon BE range, so a high-BE C~1s component is compatible with a reconstructed Ti vacancy but is not specific to it in a carbon-rich or nanocomposite sample. Subtraction of a nominal amorphous-carbon peak cannot isolate the reconstruction because interface and amorphous environments form a distribution rather than one transferable component.

Experimental nc-TiC/a-C(:H) spectra provide a direct comparison for these calculated groups. Sou{\v{c}}ek et al.\ fitted a carbon-rich nc-TiC/a-C:H coating with C--Ti, intermediate C--Ti$^*$, and C--C components centred near 282, 283, and \SI{285}{\electronvolt}, respectively \cite{Soucek2017}. Non-destructive high-kinetic-energy XPS by Lewin et al.\ placed C--Ti at \SIrange{281.5}{281.6}{\electronvolt}, C--Ti$^*$ at \SIrange{282.5}{282.6}{\electronvolt}, and two amorphous-C components within \SIrange{283.8}{284.9}{\electronvolt} \cite{Lewin2008}. The experimental C--Ti$^*$ separation from C--Ti is therefore approximately \SI{+1.0}{\electronvolt} and falls close to the calculated \SI{+0.90}{\electronvolt} mean for the C--C-bonded atoms in the terminating TiC(001) planes. The calculated amorphous-carbon mean lies below the experimental amorphous-C components, which occur \SIrange{+2.3}{+3.3}{\electronvolt} above C--Ti in the spectra of Lewin et al. The remaining difference is readily affected by the sp$^2$/sp$^3$ ratio, density, and hydrogen content of amorphous carbon. In addition, the arithmetic mean of sampled, unbroadened sites need not coincide with the maximum of a fitted experimental component.

Analysis of the fitted components in Fig.~8 of Sou{\v{c}}ek et al.\ gives approximate full widths at half maximum (FWHM) of \SI{0.60}{\electronvolt}, \SI{0.73}{\electronvolt}, and \SI{1.77}{\electronvolt} for C--Ti, C--Ti$^*$, and C--C, respectively \cite{Soucek2017}. Laboratory XPS measurements give comparable C--Ti widths of \SI{0.59}{\electronvolt} for cubic TiC and \SI{0.70}{\electronvolt} for a magnetron-sputtered TiC reference film \cite{Naslund2020,Osinger2022}. Neither Sou{\v{c}}ek et al.\ nor Lewin et al.\ required a separate component below the TiC peak. The calculated \SI{-0.30}{\electronvolt} mean for unbonded fivefold C is only about half the width of the main TiC line. At modest abundance, these sites would broaden its low-BE side rather than form a resolved peak. The simple fixed-cell construction and rapid quench of a single amorphous realization may under-cross-link the TiC termination and overrepresent these sites. The calculated BEs characterize the explicit local environments, while their frequencies remain model specific.

Beyond spectral overlap and linewidth, experimental visibility depends on the defect concentration and the number of strongly shifted atoms per vacancy. In Al-capped TiN, Ti-vacancy concentrations as low as \SI{1.2}{\percent} of the metal sublattice were quantified by XPS, aided by six shifted N neighbours per vacancy and suppression of surface oxidation \cite{Ondracka2025}. In the TiC reconstructions studied here, one or two C atoms per vacancy contribute to the highest-shift group. For dilute reconstructed vacancies in near-stoichiometric TiC, equal C~1s weights therefore give approximately \SIrange{1}{2}{\percent} of the carbide signal at \SIrange{2.3}{2.8}{\electronvolt} above the bulk TiC component for a \SI{1}{\percent} metal-sublattice vacancy concentration. This separation favours detection in a clean crystalline sample, whereas free or adventitious carbon can obscure the same spectral region \cite{GreczynskiCarbon2018}. Such a feature would be consistent with C--C-bonded Ti-vacancy reconstructions, with the overlapping shifts among the four motifs favouring assignment to reconstructed vacancy environments as a group rather than to a specific atomic configuration.

\section{Conclusions}

Self-consistent core-hole calculations predict a clear local spectroscopic distinction between an unreconstructed Ti vacancy and C--C-bonded vacancy reconstructions in TiC. The unreconstructed vacancy gives a nearest-neighbour C 1s BE shift of only \SI{-0.36}{\electronvolt}. Comparable negative BE shifts occur for unbonded, lattice-like C atoms missing one Ti neighbour in both the crystalline and interface models, consistent with a local electrostatic (Madelung-like) contribution. The strongest positive shift in each reconstructed model lies between \SI{+2.34}{\electronvolt} and \SI{+2.80}{\electronvolt}. These shifts occur at specific carbon sites in the newly formed C--C-bonded networks and provide candidate fingerprints of Ti-vacancy reconstruction in an otherwise well-ordered crystal.

In the exploratory TiC/amorphous-C model, the large BE shifts of the two central carbon atoms in the reconstructed C--C network persist but overlap the amorphous-C contribution. The \SI{+0.90}{\electronvolt} mean shift of the C--C-bonded atoms in the terminating TiC(001) planes is close to the approximately \SI{+1.0}{\electronvolt} separation of the experimental intermediate C--Ti$^*$ component from bulk TiC, supporting an interface contribution to this feature. Experimental visibility also depends on site multiplicity, vacancy abundance, and spectral broadening. The calculated energies provide reference values for experimental searches for reconstructed vacancies, while assignment of an individual motif requires information beyond peak position.

\section*{Data availability}

The DFT results supporting this work are available in the NOMAD Repository under DOI \href{https://doi.org/10.17172/nomad.jzw1-4yya}{10.17172/nomad.jzw1-4yya} \cite{OndrackaDataset2026}. NOMAD is described in Ref.~\cite{Scheidgen2023}. The full reproducible analysis workflow is available in Ref.~\cite{OndrackaWorkflow2026}.

\section*{CRediT authorship contribution statement}

\textbf{Jakub Koch:} Investigation, Data curation, Visualization, Writing -- original draft.
\textbf{Pavel Ondra{\v{c}}ka:} Conceptualization, Methodology, Supervision, Visualization, Writing -- original draft, review and editing.

\section*{Declaration of competing interest}

The authors declare that they have no known competing financial interests or personal relationships that could have appeared to influence the work reported in this paper.

\section*{Acknowledgements}

This work was supported by the Ministry of Education, Youth and Sports of the Czech Republic through e-INFRA CZ (ID:90254) and project LM2023039.

\section*{Declaration of generative AI and AI-assisted technologies in the manuscript preparation process}

During preparation of this work, the authors used OpenAI Codex to assist with literature searches, language editing, and scripting. The authors reviewed and edited the output and take full responsibility for the content of the article.

\printbibliography

\end{document}